\documentclass[preprint,12pt]{elsarticle}

\usepackage{amssymb}
\usepackage{amsmath}
\usepackage{subcaption}

\newcommand \B{\mathcal{B}}

\journal{Nuclear Physics B}

\begin{document}

\begin{frontmatter}




\title{Constraints on the Higgs boson anomalous FCNC interactions with light quarks}


\author[label1]{M.~Ilyushin}
\author[label1,label2]{P.~Mandrik}
\author[label1]{S.~Slabospitsky}

\address[label1]{NRC ``Kurchatov Institute'' - IHEP, Protvino, Moscow Region, Russia}
\address[label2]{Moscow Institute of Physics and Technology, Dolgoprudny, Moscow Region, Russia}

\address{}

\begin{abstract}
We consider the Higgs boson anomalous FCNC interactions with $u$, $c$, $d$, $s$ and $b$ quarks using the effective field theory framework. 
Constraints on anomalous couplings are derived from experimental results on Higgs boson production with subsequent decay into $b \bar{b}$ pair at LHC with $\sqrt{s} = 13 $~TeV. 
Upper limits on the branching fractions of $H \to b\bar{s}$ and $H \to b\bar{d}$ are set 
by performing a realistic detector simulation and accurately reproducing analysis selections of the CMS Higgs boson measurement in the four-lepton final state at $\sqrt{s} = 13$ TeV.
The searches are projected into operation conditions of HL-LHC.
Sensitivity at FCC-hh to anomalous FCNC interactions is studied based on Higgs boson production with $H \rightarrow \gamma \gamma$ decay channel.
It is shown that at FCC-hh machine one can expect to set the upper limits of the order of $10^{-2}$ at $95\%$ CL for $\mathcal{B}(H \rightarrow b\bar{s})$ and $\mathcal{B}(H \rightarrow b\bar{d})$.
\end{abstract}


\end{frontmatter}


\section{Introduction}
\label{section_intro}

The discovery of Higgs boson by the Large Hadron Collider
(LHC)~\cite{Aad:2012tfa, Chatrchyan:2012xdj} experiments has opened up
new area of direct searches for physics Beyond Standard Model (BSM).
One of the possible anomalous interaction is the
Higgs-mediated flavour-changing neutral currents (FCNC).
These processes are
forbidden in Standard Model (SM) at tree level and are strongly
suppressed in loop corrections by the Glashow-Iliopoulos-Maiani
mechanism~\cite{PhysRevD.2.1285}.

The Higgs mediated FCNC in top-quark sector
is actively investigated at LHC~\cite{Aad:2015pja, Aaboud:2017mfd, Aaboud:2018pob, Khachatryan:2016atv, Sirunyan:2017uae} by searching for
 $t\bar{t}$ production 
with one top quark decay through a FCNC channel and other follow
the dominant SM decay $t \rightarrow bW$.
The results of the searches are summarized in
Table~\ref{table:top_fcnc_results}.

\begin{table} [htbp]
  \centering
  \caption{The current experimental upper limits on FCNC decays of
    top-quark at 95\% CL.}\label{table:top_fcnc_results}%
  \renewcommand{\arraystretch}{1.5}
  \begin{tabular}{ c | c | c | c }
  \textbf{Detector} & $\mathcal{B}(t \rightarrow u H)$ & $\mathcal{B}(t \rightarrow c H)$ & Ref. \\ \hline
  ATLAS, 13 TeV, 36.1 fb$^{-1}$ & $1.9 \times 10^{-3}$ & $1.6 \times 10^{-3}$ & \cite{Aaboud:2018pob} \\ \
  CMS,   13 TeV, 35.9 fb$^{-1}$ & $4.7 \times 10^{-3}$ & $4.7 \times 10^{-3}$ & \cite{Sirunyan:2017uae} \\
  \end{tabular}
\end{table}

The FCNC couplings of the Higgs to the rest SM quarks can affect various
low-energy precision measurements.
The strongest indirect bounds on FCNC quark-quark-Higgs
couplings came from measurement of
$B_{d,s} - \bar{B}_{d,s}$, $K^0 - \bar{K}^0$ and $D^0 - \bar{D}^0$
oscillations~\cite{Harnik:2012pb}. The corresponding constraints on FCNC
couplings translated into upper limits on branching fractions of the
FCNC decays of Higgs boson to $u,d,s,c,b$ quarks are
summarized in the  Table~\ref{table:light_fcnc_results}.
Due to huge QCD background the experiments at LHC are less sensitive
to searching for FCNC decays of the Higgs boson.
On the other hand 
the direct probes of such processes could complement the indirect limits.
In addition in possible BSM scenarios the 
branching ratio of $H \rightarrow q q'$
can be enhanced with  keeping other low-energy flavour observables
approximately at their SM values~\cite{Crivellin:2017upt, Altmannshofer:2019ogm}.
Therefor, the searches for FCNC Higgs boson
interactions are very important and could be considered
as a complementary probe of new physics.

At the moment there is no any experimental evidence of the FCNC process.
Future research and increase of the experimental sensitivity are
related to the proposed energy-frontier colliders \cite{Mandrik:2018gud, Barducci:2017ioq, Mandrik:2018yhe, Arroyo-Urena:2019qhl}
such as High Luminosity LHC (HL-LHC) \cite{Apollinari:2116337} and Future Circular Collider (FCC-hh) project, 
defined by the target of 100 TeV proton-proton collisions with a
total integrated luminosity of 30 ab$^{-1}$~\cite{Benedikt:2651300,Mangano:2651294}.


\begin{table} [htbp]
  \centering
  \caption{
    The upper limits on FCNC decays of Higgs boson to the light quarks
    at 95\% CL from experiments with mesons
    (see \cite{Harnik:2012pb} for details).
  }\label{table:light_fcnc_results}%
  \renewcommand{\arraystretch}{1.5}
  \begin{tabular}{ c | c }
  \textbf{Observable}   & \textbf{Constraint} \\ \hline
  $D^0$ oscillations    & $\mathcal{B}(H \rightarrow u \bar{c}) \lesssim 2 \times 10^{-5}$ \\ 
  $B^0_d$ oscillations  & $\mathcal{B}(H \rightarrow d \bar{b}) \lesssim 8 \times 10^{-5}$ \\ 
  $K^0$ oscillations    & $\mathcal{B}(H \rightarrow d \bar{s}) \lesssim 2 \times 10^{-6}$ \\ 
  $B^0_s$ oscillations  & $\mathcal{B}(H \rightarrow s \bar{b}) \lesssim 7 \times 10^{-3}$ \\ 
  \end{tabular}
\end{table}

\begin{figure}
  \begin{center}
  \begin{subfigure}[t]{0.32\textwidth}
    \centering
    \includegraphics[height=3cm,clip]{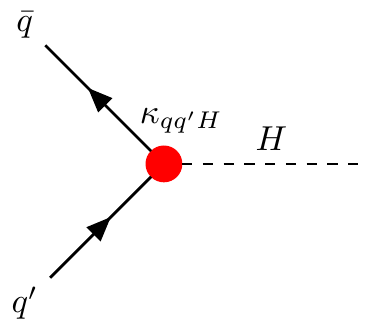}
  \end{subfigure}
  \begin{subfigure}[t]{0.32\textwidth}
    \centering
    \includegraphics[height=3cm,clip]{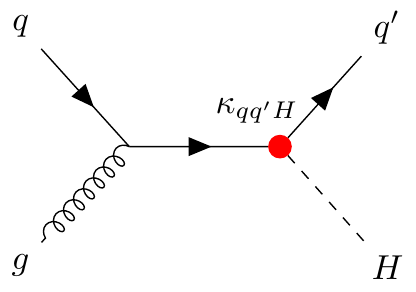}
  \end{subfigure}
  \begin{subfigure}[t]{0.32\textwidth}
    \centering
    \includegraphics[height=3cm,clip]{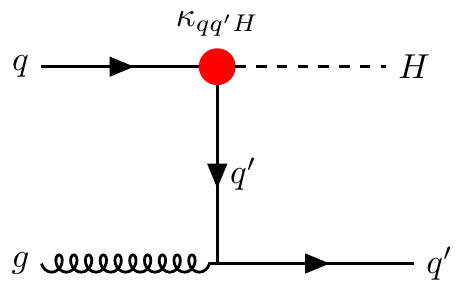}
  \end{subfigure}
  \end{center}
  
  \caption{Example of diagrams for Higgs boson production (left) and Higgs boson
    associated production with quark (center and right) mediated by FCNC
    couplings.}
  \label{feyman_fcnc_0}       
\end{figure}

In this article we invested the contribution of FCNC interactions to
the single Higgs boson production (fig.~\ref{feyman_fcnc_0}, left) 
and Higgs boson production in association with a light
quark (fig. \ref{feyman_fcnc_0}, center and right).
The limits on Higgs boson FCNC interactions based on recent LHC
data are obtained and the searches are projected into operation
conditions of HL-LHC \cite{Apollinari:2116337} and FCC-hh projects.
The cross section ratio for the different processes are presented in table 1. 

\section{The constraints from the current Higgs production cross-sections}
\label{constraints}
The flavor-violating couplings may arise from different
sources~\cite{Agashe:2013hma}. In this article we use the
effective field theory approach
(EFT)~\cite{Weinberg:1978kz, Buchmuller:1985jz, Arzt:1994gp}
for describing the effects of BSM physics in Higgs interactions.
The effective Lagrangian (up to dimension-six
gauge-invariant effective operators) has the form as 
follows~\cite{AguilarSaavedra:2004wm, AguilarSaavedra:2009mx}:
\begin{equation} \label{eq_lagrangian}
  \mathcal{L}_{BSM} = -\frac{ 1 }{ \sqrt{2} } \bar{q}
  (\kappa_{qq'H}^L P_L + \kappa_{qq'H}^R P_R) q' H
\end{equation}
where $P_{L,R} = \frac{1}{2}(1 \pm \gamma^5)$, $q, q' \in (u,c,t)$ or
$q, q' \in (d,s,b)$. The couplings  
$\kappa_{qq'H}^L$ and $\kappa_{qq'H}^R$ are complex in general.

Note, that in our analysis these couplings are appeared
in the combination
\begin{eqnarray*}
|\kappa^L_{qq'}|^2 + |\kappa^R_{qq'}|^2 = 
(\hbox{Re}\,\kappa^L_{qq'})^2 + (\hbox{Im}\,\kappa^L_{qq'})^2 +
(\hbox{Re}\,\kappa^R_{qq'})^2 + (\hbox{Im}\,\kappa^R_{qq'})^2
\end{eqnarray*}
Thus, in what follows  we set
\begin{eqnarray}
  \left.
  \begin{array}{l}
    \kappa \equiv  |\kappa^L_{qq'}| = |\kappa^R_{qq'}| 
    \\
 \lambda \equiv |\hbox{Re}\,\kappa^L_{qq'}| = |\hbox{Im}\,\kappa^L_{qq'}| = 
 |\hbox{Re}\,\kappa^R_{qq'}| = |\hbox{Im}\,\kappa^R_{qq'}| \\
  \to \kappa = \sqrt{2} \lambda
  \end{array}
  \right\}   
    \label{coupling1}
\end{eqnarray}
The Higgs decays width resulted from~(\ref{eq_lagrangian}) equals:
\begin{eqnarray}
  \Gamma(H \to q \bar{q}') =
  \frac{3 ( |\kappa^L_{qq'}|^2 + |\kappa^R_{qq'}|^2) M_H}{32 \pi} 
= \frac{3 |\kappa|^2 M_H}{16 \pi}
  = |\lambda|^2 \times 14.92 \;\;\hbox{GeV}
  \label{wid1}
\end{eqnarray}
The very rough estimates of the coupling  $\kappa_{qq'}$
could be obtained from the Higgs production  
 in the
$pp$-collisions at LHC~\cite{Aaboud:2018zhk, Sirunyan:2018kst}:
\begin{eqnarray}
  pp \, \to \, H \, X, \;\; pp \, \to \, H \; W/Z X, \quad H \to b \bar{b}
   \label{reaction0}
\end{eqnarray}
We use the experimental results from ATLAS and CMS collaborations:
\begin{eqnarray}
\mu_b = \displaystyle \frac{\sigma^{exp}(p p \to \, H \, X) }
{\sigma^{theor}(p p \to \, H \, X) }
\label{rat1}
\end{eqnarray}

\begin{center}
\renewcommand{\arraystretch}{1.2}

\begin{tabular}{l|c|c|l}
  &  $p p \to \, H \; W/Z \, X$ &  $p p \to \, H \, X$
& 
  \\ \hline
  ATLAS & $ \mu_b = 0.98^{+ 0.22}_{-0.21}$ & $\mu_b = 1.01 \pm 0.20$ &
  \cite{Aaboud:2018zhk}
  \\ \hline
  CMS &
  $\mu_b = 1.01 \pm 0.22$ & $\mu_b =
  1.04 \pm 0.20$ & \cite{Sirunyan:2018kst} 
\end{tabular}
\renewcommand{\arraystretch}{1.5}
\end{center}
\noindent 
and for estimates we set 
\begin{eqnarray}
  0.8 \leq \mu_b \leq 1.2 \label{rat2}
\end{eqnarray}
In order to get the constraints on anomalous  constants $\kappa_{qq'}$
we consider the ratio:
\begin{eqnarray}
  && \tilde{\mu}_b =
  \frac{\sigma(pp \to H)_{SM+FCNC} \, \B_{det}(H \to b \bar{b})_{SM+FCNC} }
   {\sigma(pp \to H)_{SM} \, \B_{det}(H \to b \bar{b})_{SM }} 
   \label{ratio2}
\end{eqnarray}
where $(...)_{SM}$ and $(...)_{SM+FCNC}$ stands for SM and SM+FCNC
contributions to  Higgs production and decays.
The value $\B_{det}$ equals branching fractions of the Higgs decays
into quark-antiquark pair times the  
 $B$-tagging and $B$ miss-tagging efficiencies
(from ATLAS paper~\cite{Aaboud:2018zhk})
\begin{eqnarray}
  \varepsilon_b = 70\%, \;
 \varepsilon_c = 12\%, \;
 \varepsilon_{q} = 0.3\%, \; q = d,u,s \label{btag}
 \end{eqnarray}
So, for SM and SM+FCNC scenarios we have:
\begin{eqnarray*}
   \B_{det}(H \to b \bar{b})_{SM } &=& \B_{sm}(H\to b \bar{b})
   \varepsilon_b^2 \\
   \B_{det}(H \to b \bar{b})_{SM+FCNC} & =& 
 B_{fcnc}(H\to b \bar{b})  \varepsilon_b^2 +
 B_{fcnc}(H \to q_1 \bar{q}_2) \varepsilon_{q_1}  \varepsilon_{q_2}
 \end{eqnarray*}
We use the {\scshape MG5\_}a{\scshape MC@NLO}~2.5.2~\cite{Alwall:2014hca}
package (see section \ref{Event_generation}) for estimation of the Higgs anomalous production
cross-sections at $\sqrt{s} = 13$~TeV: 
\begin{eqnarray*}
   \sigma_{sm} \approx 50\, \hbox{pb} && 
  \sigma(b \bar{s} + \bar{b} s)_{fcnc} =
  |\lambda|^2 \times 18000 \;\; \hbox{pb} \\ 
  \sigma(b \bar{d} + \bar{b} d)_{anom} =
  |\lambda|^2 \times 45600 \; \hbox{pb}  
  && \sigma(c \bar{u} + \bar{c} u)_{anom} =  |\lambda|^2 \times 82000 \;
  \hbox{pb}
\end{eqnarray*}
Then, from requirement on $\tilde{\mu}_b$ from~(\ref{rat2}) 
we get the constraints on the anomalous couplings $\kappa_{q q'}$.
To avoid ambiguities due to different normalizations of the couplings in the Lagrangian, the
branching ratios of the corresponding FCNC processes are also used for presentation of the results.

\begin{table}[h]
\centering
\caption{The upper limits on the anomalous couplings, the Higgs
  boson decay widths (in MeV) and branching fractions. 
}
      \label{fcnc_xsec_table_22}
      \vspace{5pt}
      \renewcommand{\arraystretch}{1.5}
\begin{tabular}{c|c|c|c|c} \hline \hline
  $q q'$  & $\kappa$        & $\lambda$  & $\Gamma(q \bar{q}')$~MeV & $\mathcal{B}(q \bar{q}')$ \\  \hline
  $bs$    & $ \leq 0.0085$  & $0.006$    & 0.54                     & 10\%  \\ 
  $bd$    & $  \leq 0.0089$ &  $ 0.0063$ & 0.60                     & 11\%  \\ 
  $cu$    & $ \leq 0.0096$  & $0.0068$   & 0.69                     & 13\%  \\ \hline
\end{tabular}
    \end{table}
\renewcommand{\arraystretch}{1.0}
Certainly, these constraints are much worse as indirect constraints,
given if the Table~\ref{table:light_fcnc_results}. However,
these constraints are first ones resulted from direct searches
of the Higgs FCNC interactions with the light quarks.

\section{Event generation} \label{Event_generation}
The estimation based on (\ref{ratio2}) does not take into account the differences in kinematics of the SM and FCNC Higgs boson production processes.
In order to accurately incorporate detector effects and reconstruction efficiencies for the next sections we are performing Monte-Carlo (MC) simulation of related processes.
We use the Lagrangian~(\ref{eq_lagrangian}) for the signal simulation.
The Lagrangian (\ref{eq_lagrangian}) is implemented in
FeynRules~\cite{Alloul:2013bka} based on~\cite{Amorim:2009mx}
and the model is interfaced with generators using the UFO
module~\cite{Degrande:2011ua}.
The events are generated using the
{\scshape MG5\_}a{\scshape MC@NLO}~2.5.2~\cite{Alwall:2014hca} package, 
with subsequent showering and hadronization in
{\scshape Pythia}~8.230~\cite{Sjostrand:2014zea}.
The {\scshape NNPDF3.0} \cite{Ball:2014uwa} PDF sets are used.
The detector simulation has been performed with the fast simulation
tool~{\scshape Delphes}~3.4.2~\cite{deFavereau2014}
using the corresponding detectors parameterization cards. No additional
pileup interactions are added to the simulation. 
The cross-sections for Higgs boson productions associated with zero or
one jet and mediated by FCNC couplings in proton-proton collisions 
for different centre-of-mass energy are given in
the Table~\ref{fcnc_xsec_table_1}. Note, these values are evaluated for
Higss production with 0 or 1 jet using the MLM matching scheme \cite{Alwall:2007fs}. Therefore, they are greater then
those used in previous section.

\begin{table}[h]
\centering
\caption{ The cross-sections of Higgs boson + 0, 1 jet productions mediated
  by FCNC couplings in proton-proton collisions for different
  centre-of-mass energies. }
      \label{fcnc_xsec_table_1}
      \vspace{5pt}
      \renewcommand{\arraystretch}{1.5}
      \begin{tabular}{c|c c c c}
      \hline
      \hline
subprocess & \multicolumn{4}{c}{ Cross section, pb } \\
  & 13 TeV & 14 TeV & 27 TeV & 100 TeV      \\ \hline
$ucH$  & $ 9.08 \times 10^{4}\lambda_{ucH}^2 $
& $ 9.85 \times 10^{4}\lambda_{ucH}^2 $  & $ 2.01 \times 10^{5}\lambda_{ucH}^2 $
& $ 7.3 \times 10^{5}\lambda_{ucH}^2 $
\\  
$dsH$  & $ 8.25 \times 10^{4}\lambda_{dsH}^2 $  &
$ 9.02 \times 10^{4}\lambda_{dsH}^2 $  & $ 1.91 \times 10^{5}\lambda_{dsH}^2 $
& $ 7.23 \times 10^{5}\lambda_{dsH}^2 $
\\  
$dbH$  & $ 4.81 \times 10^{4}\lambda_{dbH}^2 $  &
$ 5.32 \times 10^{4}\lambda_{dbH}^2 $  & $ 1.18 \times 10^{5}\lambda_{dbH}^2 $
& $ 4.77 \times 10^{5}\lambda_{dbH}^2 $
\\  
$sbH$  & $ 2.32 \times 10^{4}\lambda_{sbH}^2 $ &
$ 2.61 \times 10^{4}\lambda_{sbH}^2 $  & $ 6.67 \times 10^{4}\lambda_{sbH}^2 $
& $ 3.27 \times 10^{5}\lambda_{sbH}^2 $
\\   \hline
 \end{tabular}
 \end{table}

\section{Constrain from Higgs boson measurement in the four-lepton final state at $\sqrt{s} = 13$ TeV} \label{section_ZZ}
In this section, we obtain upper the limit on the
$\mathcal{B}(H \rightarrow b\bar{s})$ and
$\mathcal{B}(H \rightarrow b\bar{d})$ branching fractions using
constraints on Higgs boson measurement in the four-lepton final state
at $\sqrt{s} = 13$ TeV from CMS experiment at LHC~\cite{Sirunyan:2017exp}
from the reaction as follows:
\begin{eqnarray}
  p \, p \, \to \, H X, \;\; \to \, H \, j \, X, \quad
  H \to Z Z, \;\; Z \to \ell^+, \ell^-, \; \ell = \mu, e
  \label{reacHZZ}
\end{eqnarray}  
In order to accurately incorporate the effects of the analyses efficiency
different for the SM and FCNC Higgs boson production we reproduce the
events selections from~\cite{Sirunyan:2017exp}.

The four-lepton candidates build $ZZ$ pairs. One $Z$ candidate is defined
as pairs of two opposite charge and matching flavour leptons
$(e^+ e^-, \mu^+ \mu^-)$ that satisfy $12<m_{ll}<120$ GeV. Electrons are
reconstructed within the geometrical acceptance defined by pseudorapidity
$|\eta^e|<2.5$ and for transverse momentum
$p_{T}^{e}>7$ GeV. Muons are reconstructed within the geometrical acceptance
$|\eta^\mu|<2.4$
and $p_T^\mu>5$ GeV. All leptons within $ZZ$ pairs must be separated in
angular space by at least $\Delta R(l_i,l_j)>0.02$. Two of the four
selected leptons should have $p_{T,i} > 20$ GeV and $p_{T,j}>10$ GeV.

The $Z$ candidate with reconstructed mass $m_{ll}$ closest to the nominal
$Z$ boson mass is denoted as $Z_1$, and the second one is denoted as $Z_2$.
The $Z_1$ invariant mass must be larger than 40 GeV. In the $4\mu$ and $4e$
sub-channels the $ZZ$ event with reconstructed mass $m_{Z2}\ge12$ GeV and
$m_{Z1}$ closest to the nominal $Z$ boson mass. All four opposite-charge
lepton pairs that can be built with the four leptons
(irrespective of flavor) are required
to satisfy $m_{l_i^+l_j^-}>4$ GeV. Finally, the four-lepton invariant mass
should be of the Higgs boson in a $118<m_{4l}<130$ GeV.

The comparison of selection efficiencies for FCNC Higgs boson production
processes are presented in Table \ref{CMS_eff}. The simulation of the SM
Higgs boson production with Delphes show good agreement with reference
Geant4 results taken from \cite{Sirunyan:2017exp}. The selection efficiency
is different for different FCNC Higgs boson productions processes due to
the presence of the valence $d$ quark in $bdH$ vertex (as compared to
$b s \to H$ production).

    \begin{table}[h]
    \centering
    \caption{ The comparison of selection efficiency for FCNC and SM Higgs
      boson productions for different $ZZ$ decay channels before the cut on invariant mass
    reconstructed Higgs boson $m_{4l} \in [118,130]$ GeV and after the cut. The reference Geant4
      results are taken from \cite{Sirunyan:2017exp}. }
      \label{CMS_eff}
      \vspace{5pt}
      \renewcommand{\arraystretch}{1.5}
      \begin{tabular}{c|c c c c | c}
      \hline
      \hline
      Higgs production                     &    $4e$    &    $2e2\mu$   &    $4\mu$    & total   & total ($m_{4l}$ cut)  \\ \hline
      SM (Geant4)                          & 5.1\%      & 12.9\%        & 10.2\%       & 28.3\%  & 24.9\%  \\
      SM (Delphes)                         & 4.9\%      & 13.1\%        & 9.3\%        & 27.2\%  & 25.6\%  \\
      FCNC ($dbH$)                         & 3.6\%      & 9.5\%         & 6.5\%        & 19.5\%  & 17.8\%  \\
      FCNC ($sbH$)                         & 4.9\%      & 12.8\%        & 9\%          & 26.7\%  & 24.5\%  \\
      \hline
      \end{tabular}
    \end{table}

Statistical analyses is performed based on the number of selected events (after the cut on $118 < m_{4l} < 130$ GeV)
where the expected number of signal FCNC events is from our modeling and the observed and expected number of background events are taken from the CMS experimental results \cite{Sirunyan:2017exp}. 
For the signal processes lepton energy resolution (20\%), lepton energy scale (0.3\%), lepton
identification (9\% on the overall event yield) and luminosity (2.6\%)
uncertainties are taken into account.
The uncertainty from the renormalization and factorization scale is
determined by varying
these scales between 0.5 and 2 times their nominal value while keeping
their ratio between 0.5 and 2 \cite{deFlorian:2016spz}.
PDF uncertainty is determined by taking the root mean square of the
variation when using different replicas of the default PDF
set~\cite{Butterworth:2015oua}.
Contributions of the systematic uncertainties to selection efficiency of
the FCNC Higgs boson production are summarized in the
Table~\ref{CMS_sys_contribution}. The total uncertainties on the number of selected signal and background (extracted from \cite{Sirunyan:2017exp}) events are incorporated into statistical model as a nuisances neglecting the correlations.

    \begin{table}[h]
      \centering
      \caption{ Summary of contribution of the systematic uncertainties to the selection efficiency of the FCNC Higgs boson production. }
      \label{CMS_sys_contribution}
      \vspace{5pt}
      \renewcommand{\arraystretch}{1.5}
      \begin{tabular}{c|c c c c}
      \hline
      \hline
      Process                   & $\mathcal{B}(H \rightarrow b\bar{s})$ & $\mathcal{B}(H \rightarrow b\bar{d})$ \\ \hline
      Lepton energy resolution  & $< \pm 0.2\%$                         & $< \pm 0.2\%$ \\
      Lepton energy scale       & $< \pm 0.5\%$                         & $< \pm 0.5\%$ \\
      Lepton identification     & $\pm 9\%$                             & $\pm 9\%$    \\
      Luminosity                & $\pm 2.6\%$                           & $\pm 2.6\%$  \\
      QCD scale                 & $-19.6\%$ $+18.1$                     & $-17\%$ $+15.2\%$ \\
      PDF                       & $\pm 8\%$                             & $\pm 3.4$ \\ \hline
      Total                     & $-23.1\%$ $+21.9\%$                     & $-19.7\%$ $+18.2\%$ \\
      \hline
      \end{tabular}
    \end{table}

Bayesian inference is used to derive the posterior probability based on the following likelihood function:
\begin{eqnarray*}
    \mathcal{L} = \mathcal{G}\Big( N_{obs}| N_{back} + (N_{SM} + N_{FCNC}) \cdot \frac{\mathcal{B}_{FCNC + SM}}{\mathcal{B}_{SM}}, \sqrt{N_{obs}}  \Big) \times \\ 
     \times \mathcal{G}\Big( N_{back} | N_{back}^{exp}, \sigma_{N_{back}^{exp}} \Big) \times \\
     \times \mathcal{G}\Big( N_{FCNC} | N_{FCNC}^{exp}(\lambda), \sigma_{N_{FCNC}^{exp}(\lambda)} \Big)
\end{eqnarray*}
where the $\mathcal{G}$ - Gaussian function, $N_{back}^{exp},N_{SM}^{exp},N_{FCNC}^{exp}$ - the expected from the MC simulation number of background, SM and FCNC Higgs boson production events respectively, $\sigma_{N_{...}^{exp}}$ - its uncertainty, $\mathcal{B}_{FCNC + SM}$ - branching of $H \rightarrow 4\ell$ ($\ell = e, \mu$) in the presence of FCNC.

The 95\% C.L. expected exclusion limits on the anomalous couplings and the branching fractions are given in Table \ref{FCC_limits}.

\section{Sensitivity at HL-LHC}
The reconstruction efficiency estimated in section \ref{section_ZZ} can be used to project the FCNC searches into HL-LHC conditions,
defined by total integrated luminosity of 3 ab$^{-1}$ and collision energy of 14 TeV, respectively.
For the rescaling the crossections of SM Higgs boson productions are taken from \cite{Cepeda:2019klc}.
The rescaling factors for crossections of $qq \rightarrow ZZ$ and $gg \rightarrow ZZ$ background processes are taken from \cite{CMS-PAS-FTR-18-014}.
The rescaling factros for crossections of ``$Z+X$'' background processes is estimated using the corresponding crossections from {\scshape MG5\_}a{\scshape MC@NLO}~2.5.2~\cite{Alwall:2014hca} simulation of dominated $Z + jets$ process.
The cross section ratio for the different processes are summurised in table \ref{table:xsec_rescale_table}.
Statistical analyses from section \ref{section_ZZ} is reproduced for the new conditions. 
The dominated systematic uncertainties on the simulation originating from theoretical sources are scaled by 50$\%$ following the treatment of systematic uncertainties
in \cite{Cepeda:2019klc}.
In this considered scenario the theoretical uncertainties are expected to improve over time due to developments in the calculations, techniques and orders considered.
The 95\% C.L. expected exclusion limits on the anomalous couplings and the branching fractions are given in Table \ref{FCC_limits}.

\begin{table} [htbp]
  \centering
  \caption{ Cross section ratios $\sigma_{\text{14 TeV}}/\sigma_{\text{13 TeV}}$ for FCNC and background processes. }\label{table:xsec_rescale_table}%
  \renewcommand{\arraystretch}{1.5}
  \begin{tabular}{ c | c }
  \textbf{Process}      & $\sigma_{\text{14 TeV}}/\sigma_{\text{13 TeV}}$ \\ \hline  
  $qq \rightarrow ZZ$   & 1.17       \\ 
  $gg \rightarrow ZZ$   & 1.13       \\ 
  ``$Z+X$''             & 1.11       \\ 
  SM Higgs              & 1.13       \\ 
  FCNC Higgs ($dbH$)    & 1.10       \\ 
  FCNC Higgs ($sbH$)    & 1.13       \\ 
  \end{tabular}
\end{table}

\section{Sensitivity at FCC-hh}
In this section the sensitivity to single Higgs boson production through
FCNC in $bdH$ and $bsH$ subprocesses is explored for the FCC-hh experimental
conditions following the \cite{L.Borgonovi:2642471} SM study.
The $H \rightarrow \gamma \gamma$ decay channel is used in this
analysis.
The SM single Higgs production is considered as background in additional to QCD di-photon
productions including the huge tree level $qq \rightarrow \gamma \gamma$
component, generated up to two merged extra-jets, and a smaller
loop-induced component, $gg \rightarrow \gamma \gamma$, generated up to
one additional  merged jet.  A conservative K-factor of 2 is applied
to both QCD contributions.
The signal and background process generation and detector simulation
are described in \ref{Event_generation} chapter.

\begin{figure}[]
  \centering
  \includegraphics[width=0.48\columnwidth]{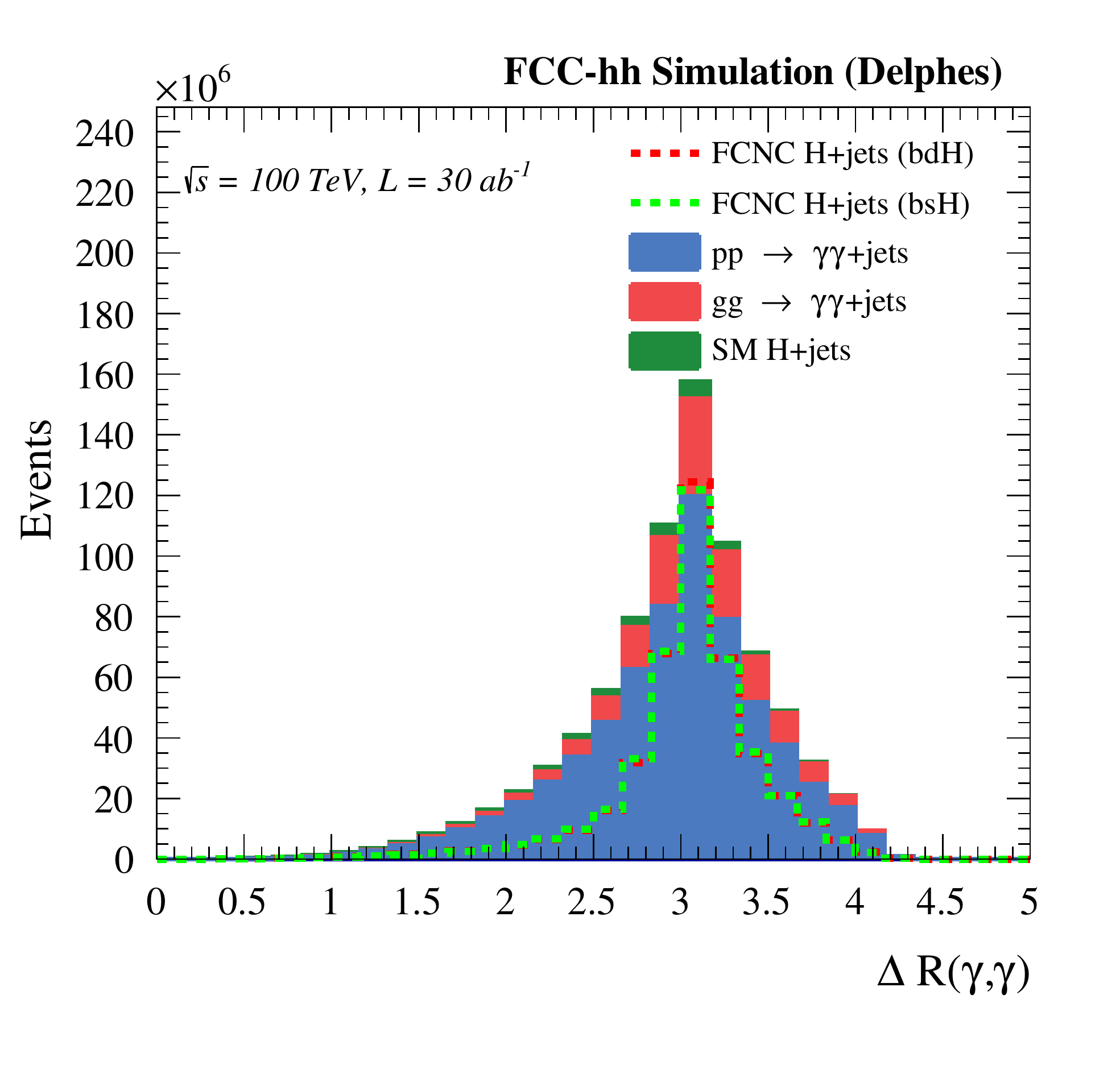}
  \includegraphics[width=0.48\columnwidth]{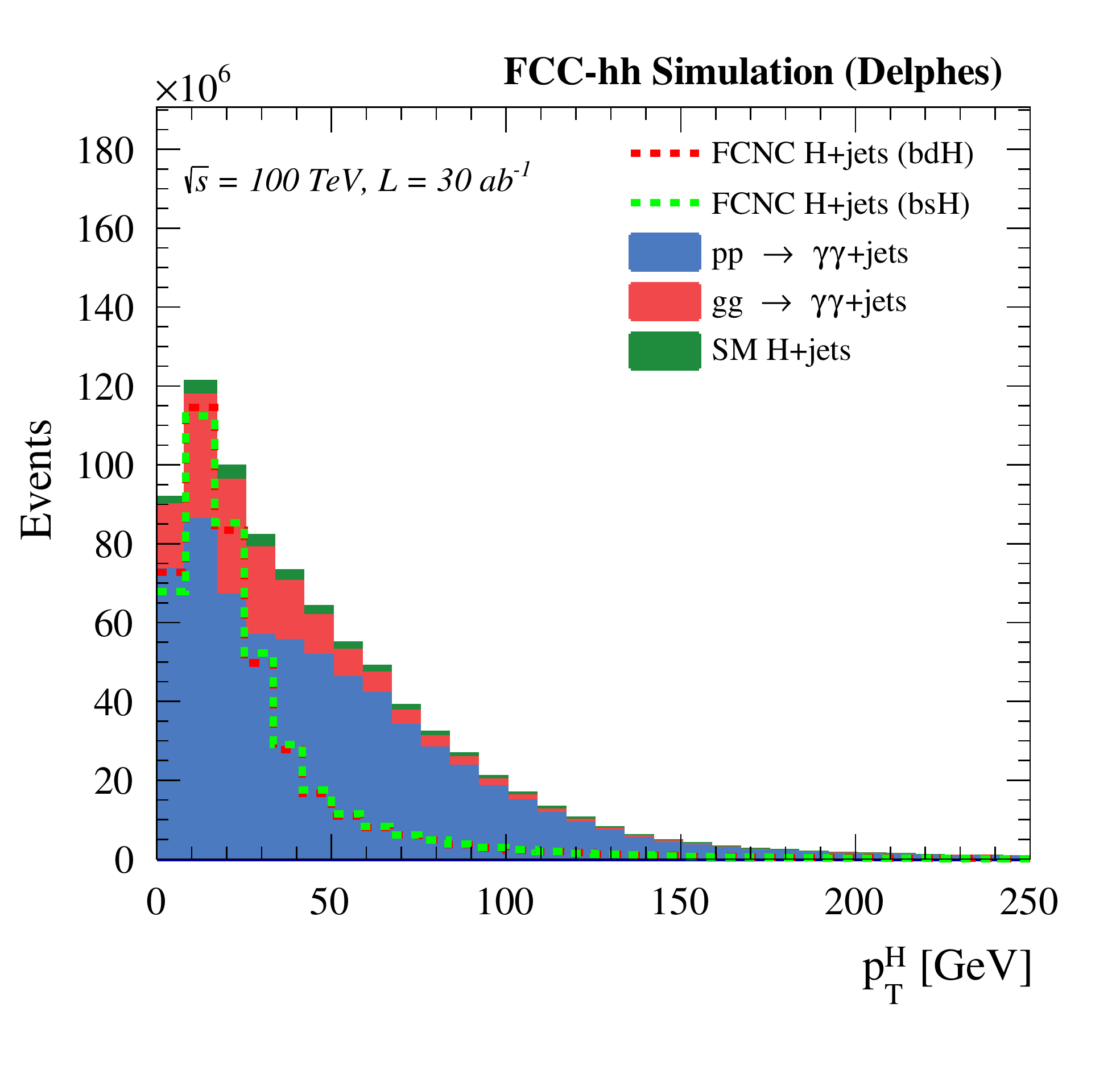} \\
  \includegraphics[width=0.48\columnwidth]{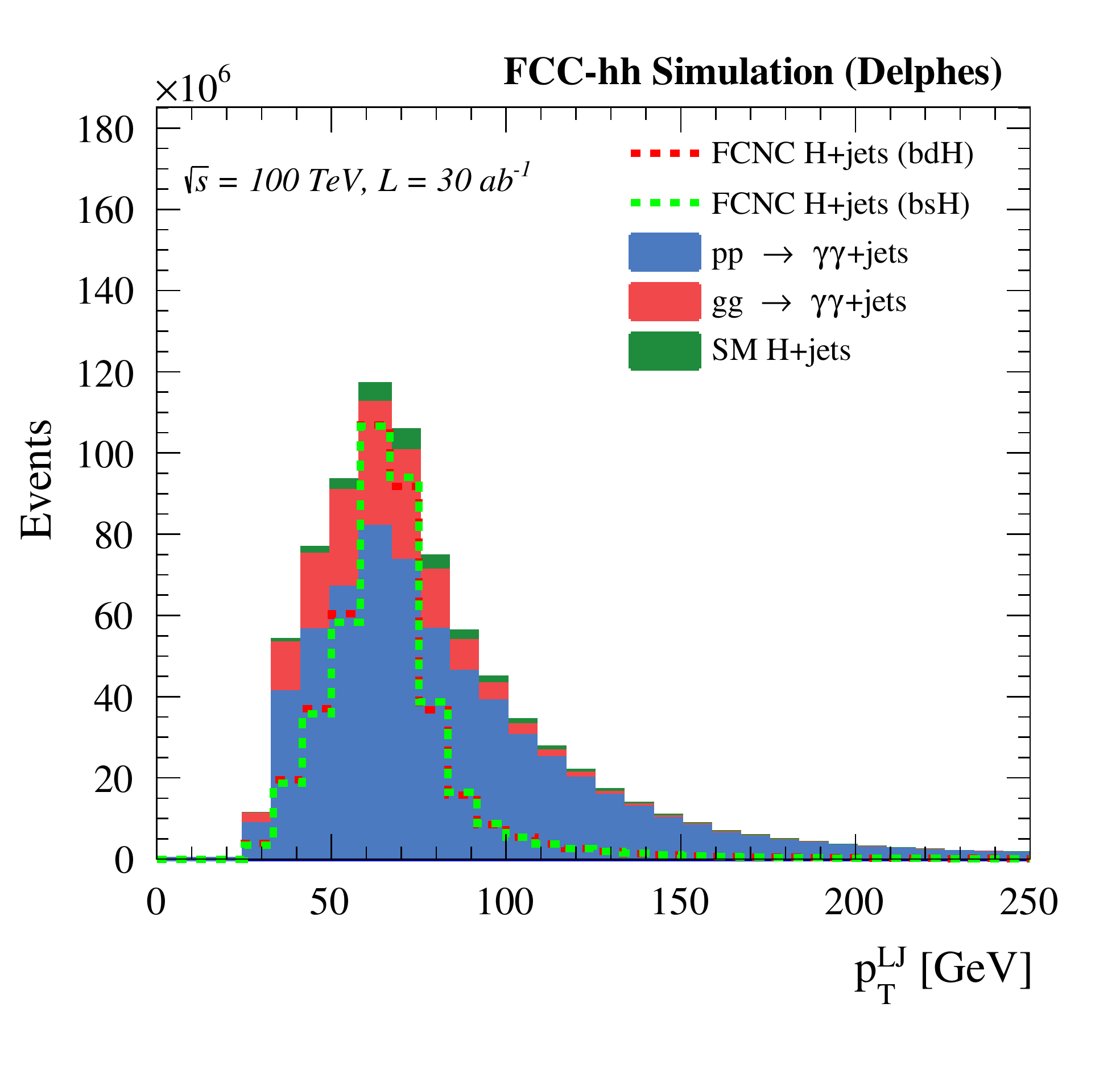}
  \includegraphics[width=0.48\columnwidth]{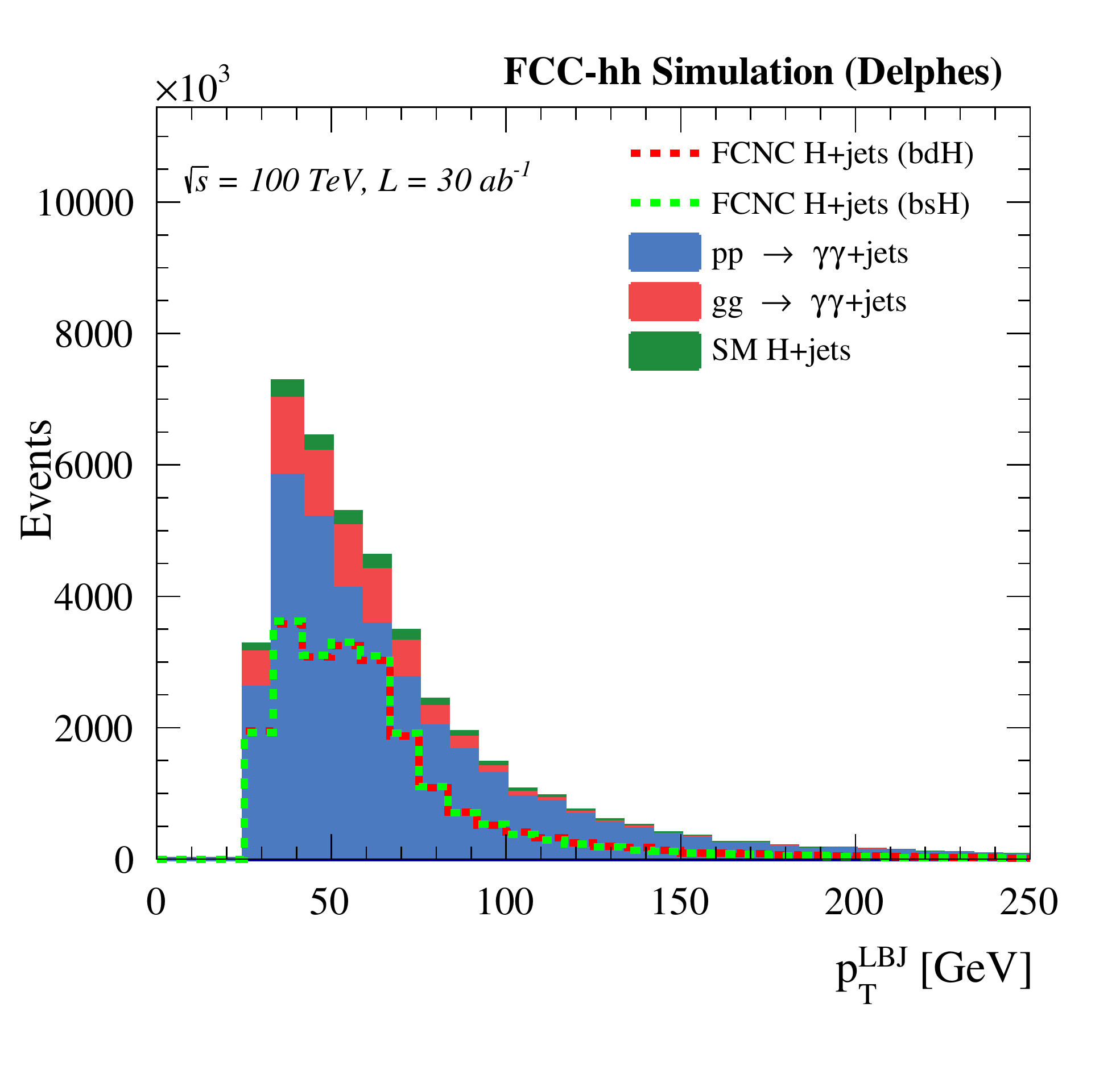} \\
  \caption{ Distributions of the kinematic variables obtained after basic selections: 
      $\Delta R$ between two selected photons with the highest $p_T$ (top-left), 
      $p_T$ of the Higgs boson candidate (top-right), 
      $p_T$ of the leading jet (bottom-left), 
      $p_T$ of the leading b-tagged jet (bottom-right). 
The signal processes have arbitrary normalization for the illustration purpose. }
  \label{fig:fcc_plots}
\end{figure}

\begin{figure}[]
  \centering
  \includegraphics[width=0.48\columnwidth]{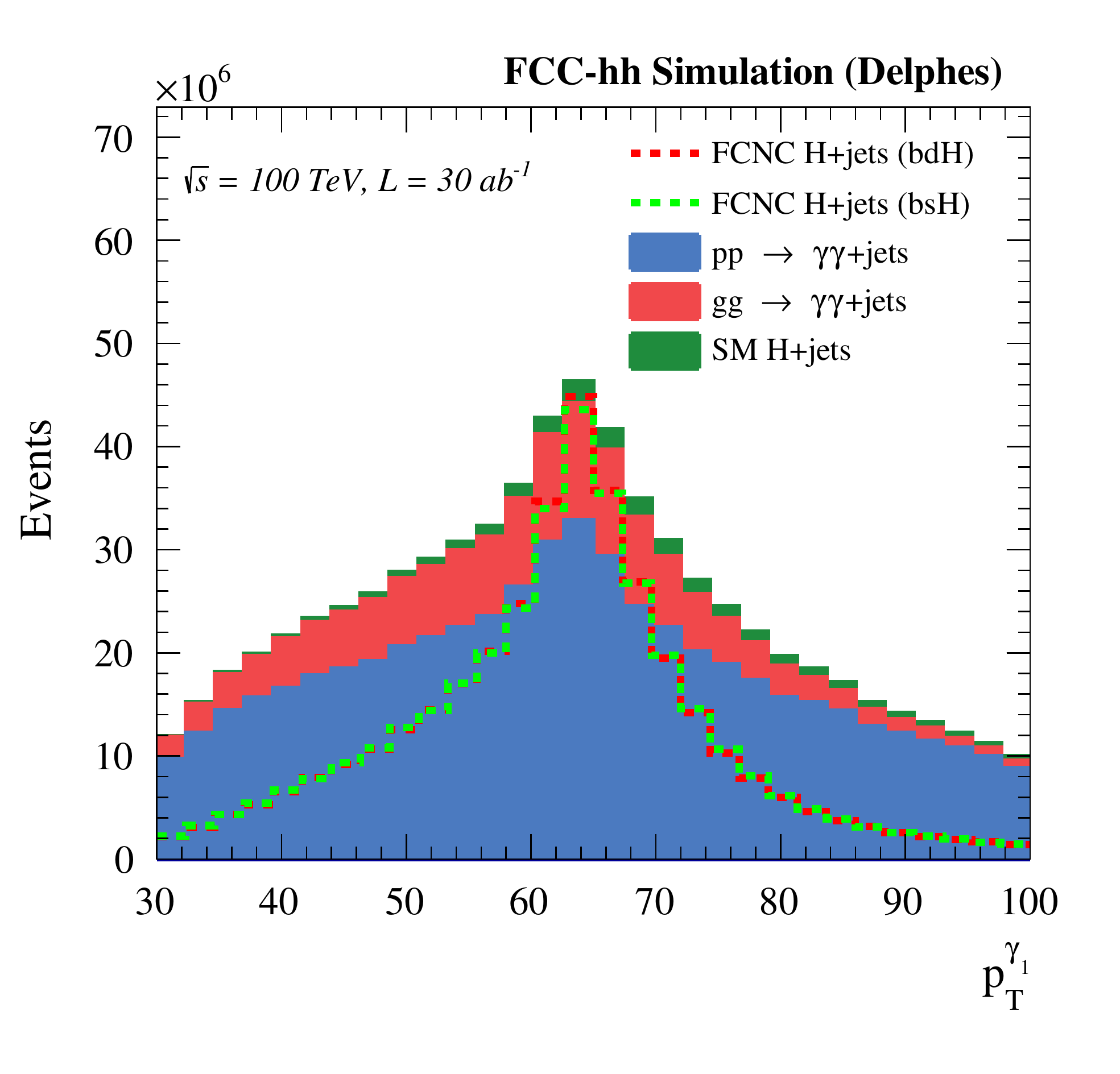}
  \includegraphics[width=0.48\columnwidth]{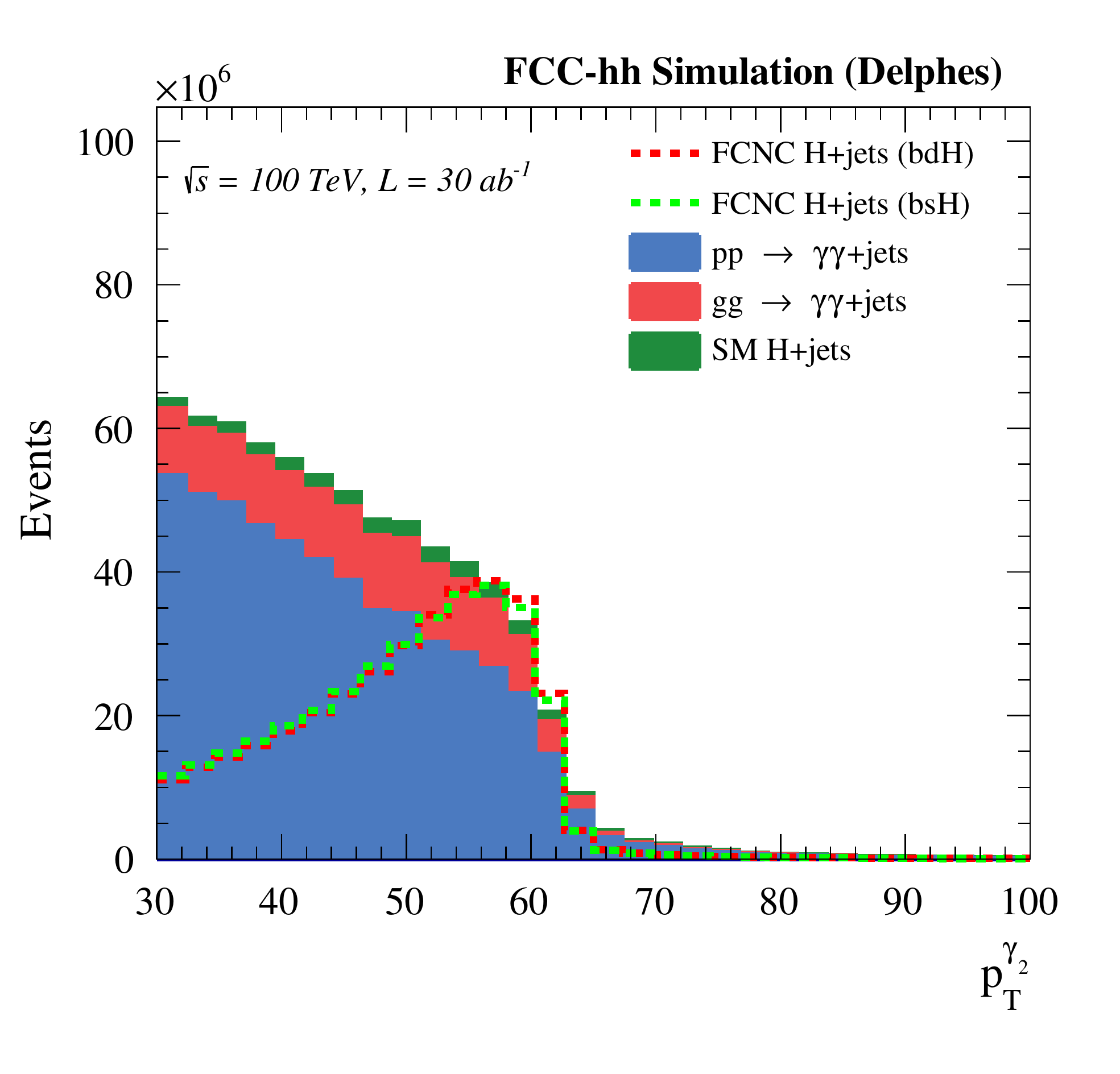} \\
  \includegraphics[width=0.48\columnwidth]{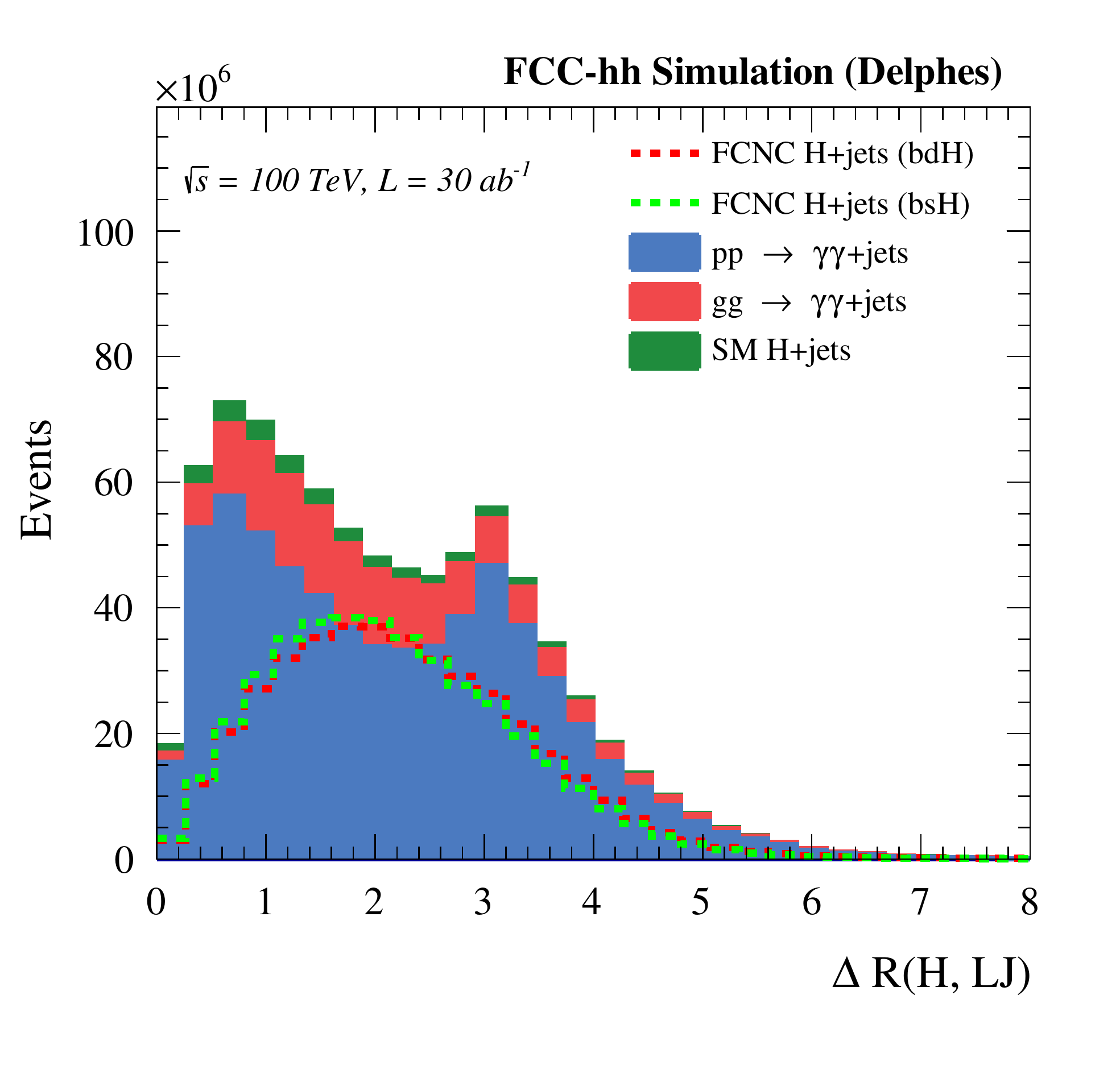}
  \includegraphics[width=0.48\columnwidth]{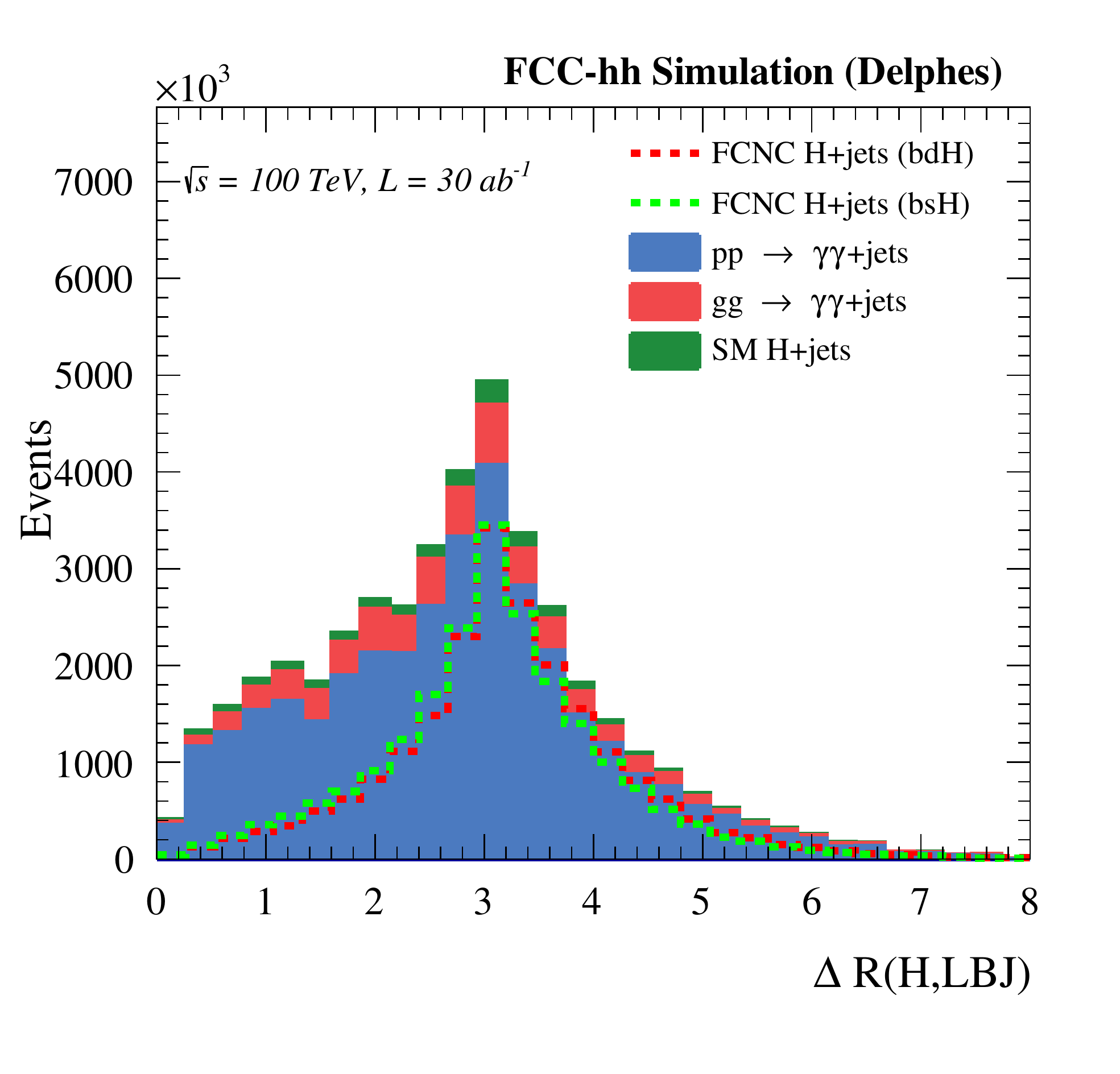} \\
  \caption{ Distributions of the kinematic variables obtained after basic selections: 
      leading photons $p_{T}^{\gamma_{1}}$ (top-left), 
      second photons $p_{T}^{\gamma_{2}}$ (top-right), 
      $\Delta R$ between Higgs boson candidate and leading jet (bottom-left), 
      $\Delta R$ between Higgs boson candidate and leading b-tagged jet (bottom-right). 
The signal processes have arbitrary normalization for the illustration purpose. }
  \label{fig:fcc_plots_2}
\end{figure}

\begin{figure}[]
  \centering
  \includegraphics[width=0.48\columnwidth]{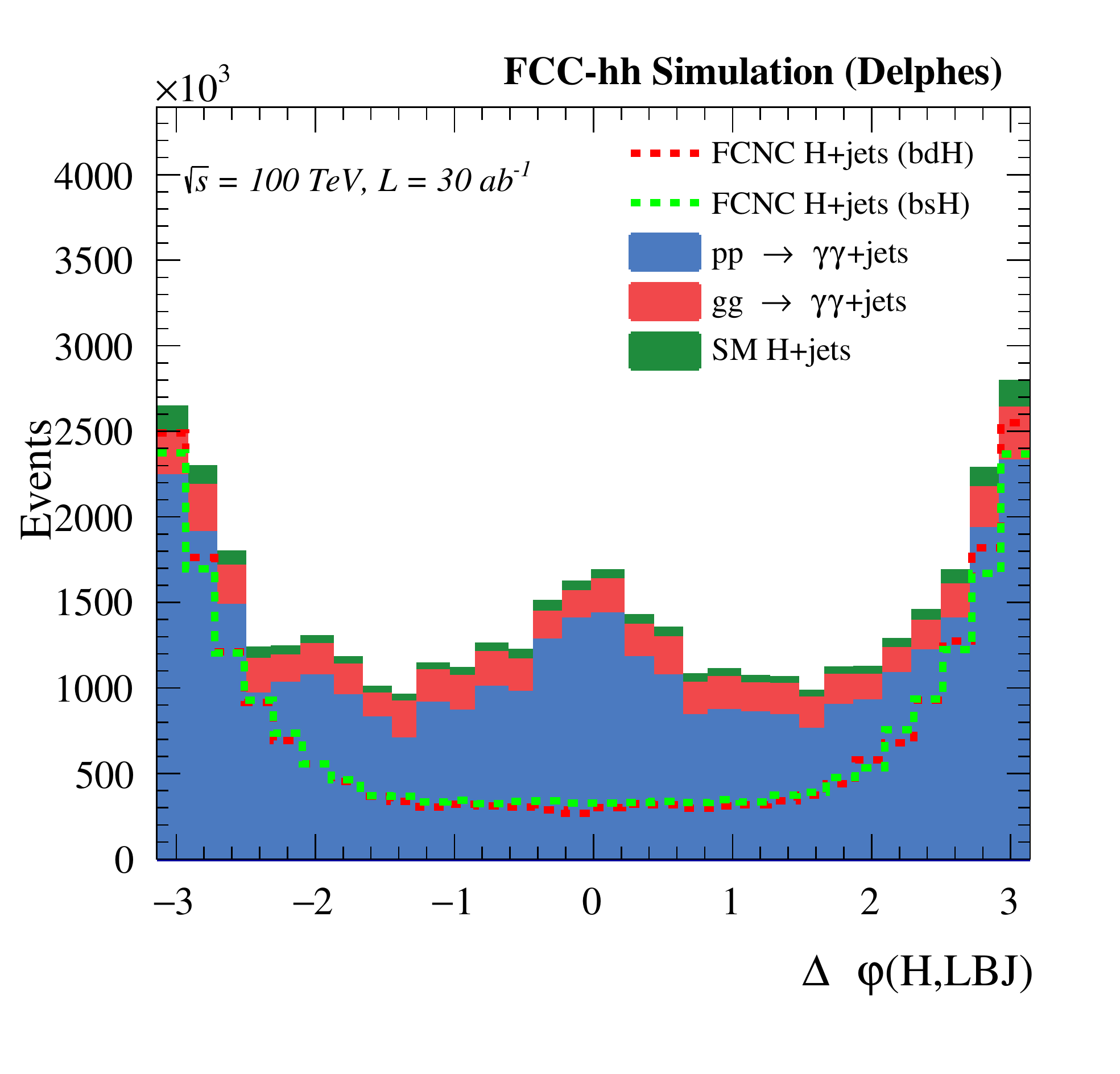}
  \includegraphics[width=0.48\columnwidth]{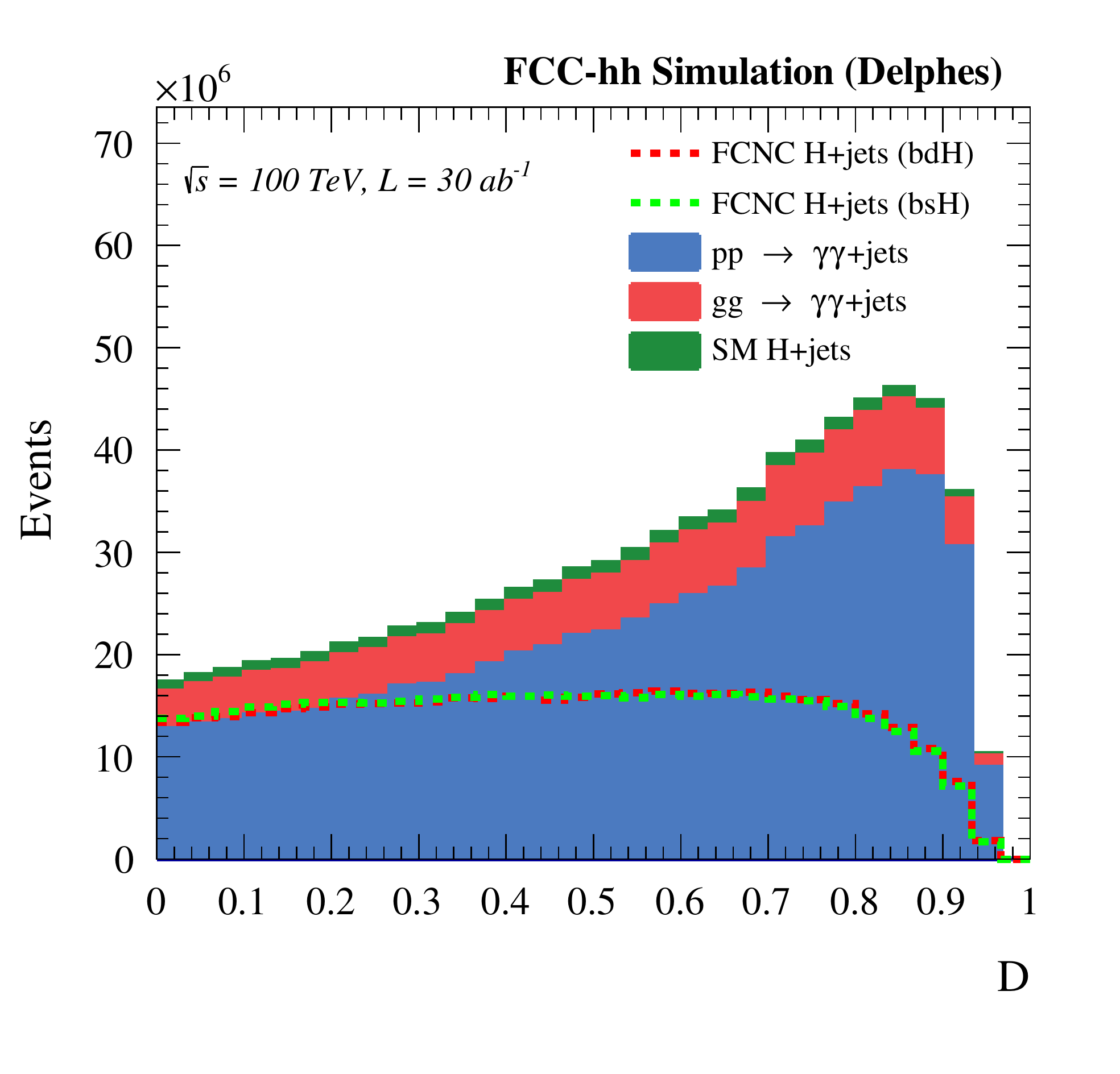} \\
  \includegraphics[width=0.48\columnwidth]{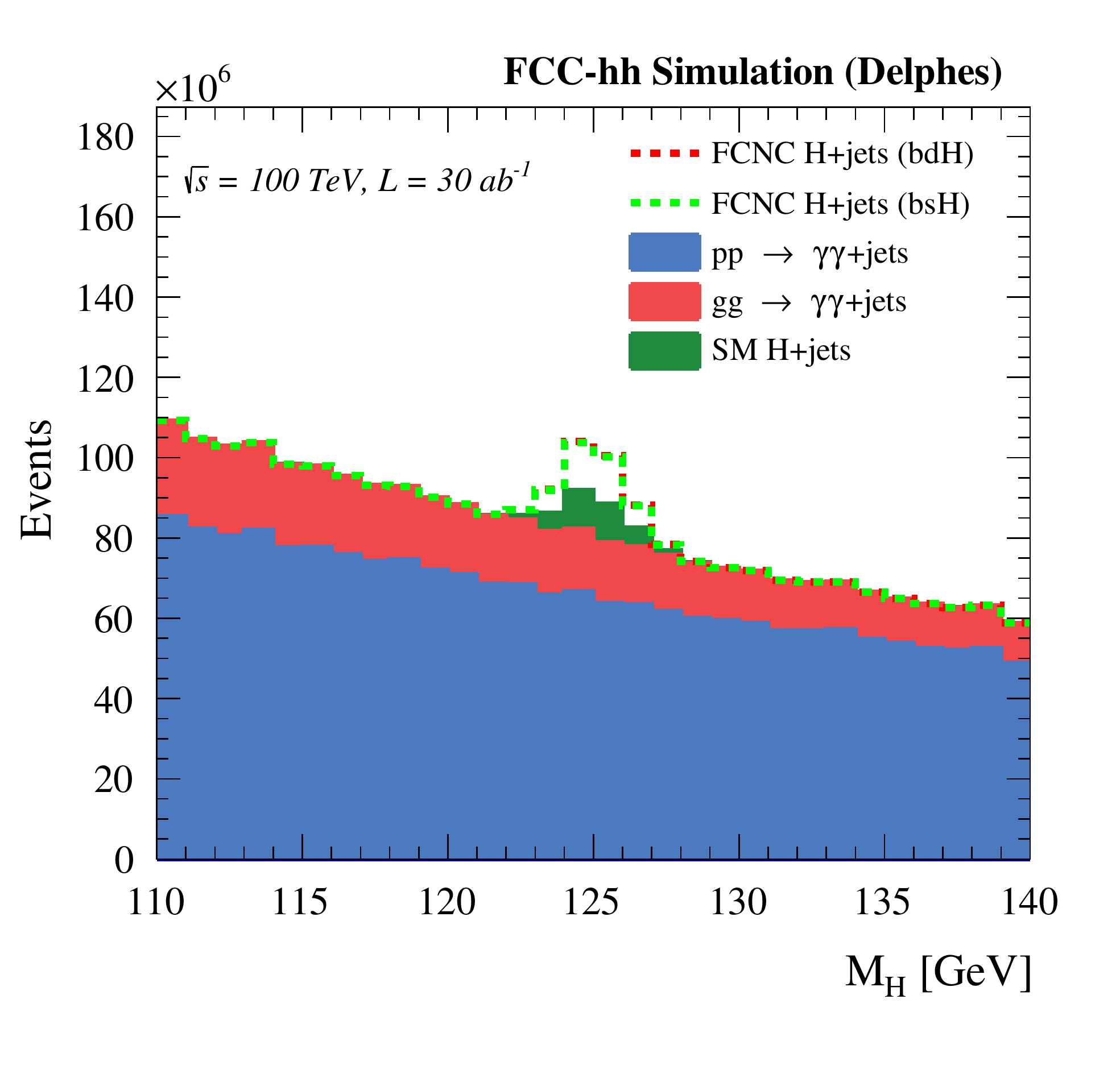}
  \includegraphics[width=0.48\columnwidth]{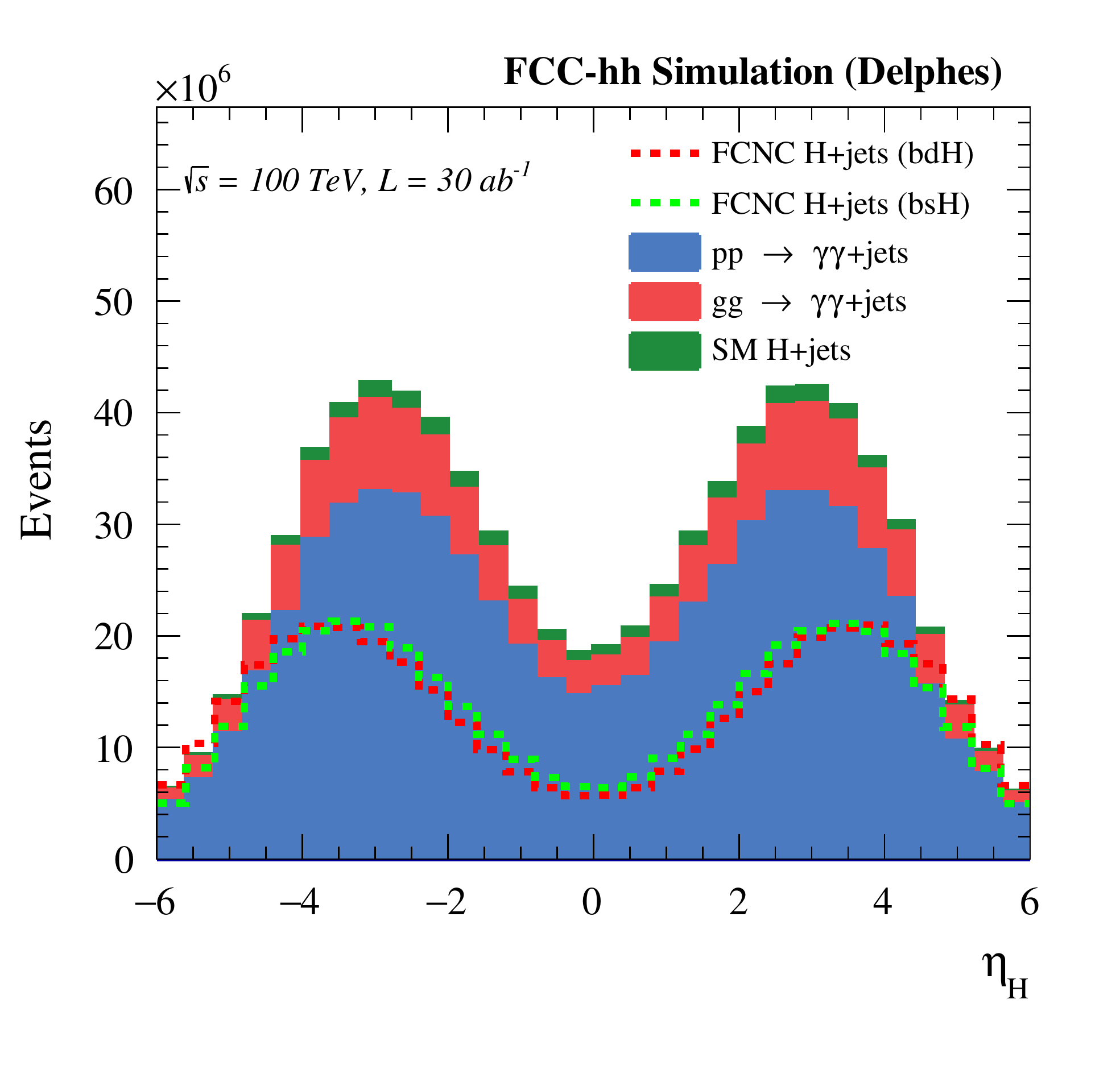}
  \caption{ Distributions of the kinematic variables obtained after basic selections: 
    $\Delta\varphi$ between reconstructed Higgs boson candidate and leading b-tagged jet (top-left), 
    disbalance in energy of photons from Higgs decay (top-right, see text for the description) and mass of the reconstructed Higgs boson candidate (bottom-left, without cut on mass), 
    $\eta$ of the reconstructed Higgs boson candidate (bottom-right). 
    The signal processes have arbitrary normalization for the illustration purpose.
  }
  \label{fig:fcc_plots_3}
\end{figure}

The photons with $p_T > 25$ GeV, $|\eta| < 4$ and relative isolation
$< 0.15$ are used in the following analyses.
Jets are reconstructed using anti$-kT$ algorithm with distance parameter
$R=0.4$ and required to have $p_T > 30$ GeV, $|\eta| < 3$.
The events are selected using the following baseline criteria:
\begin{enumerate}
\item at least 2 selected photons and at least one of them with
  $p_T > 30$ GeV;
\item mass of the Higgs boson candidate reconstructed from the two
  photons with the highest $p_T$ should be $|m_H - 125| < 5$ GeV.
\end{enumerate}
Distributions of the kinematic variables obtained after baseline selections are presented at Fig. \ref{fig:fcc_plots}, Fig. \ref{fig:fcc_plots_2} and Fig. \ref{fig:fcc_plots_3}.

A  Boosted  Decision  Tree  (BDT)  constructed  in  the  TMVA  framework  \cite{Hocker:2007ht}  is  used  to
separate  the  signal  signature  from  the  background  contributions.   10\%  of  events  selected  for
training and the remainder are used in the statistical analysis of the BDT discriminants with the
CombinedLimit package.  The following input variables are used for training:

\begin{enumerate}
  \item Higgs boson candidate $M_H$, $p_{T}^{H}$ and $\eta_{H}$;
  \item leading jet (LJ) $p_{T}^{LJ}$ and $\eta_{LJ}$;
  \item leading b-tagged jet (LBJ) $p_{T}^{LBJ}$ and $\eta_{LBJ}$;
  \item leading photons $p_{T}^{\gamma_{1}}$, $\eta^{\gamma_{1}}$ and second photons $p_{T}^{\gamma_{2}}$, $\eta^{\gamma_{2}}$;
  \item Number of jets $N_{jets}$ and number of b-tagged jets $N_{b-jets}$;
  \item $\Delta R(\gamma,\gamma)$ between leading and second photon;
  \item $\Delta R(H,LBJ)$ between Higgs boson candidate and leading jet;
  \item $\Delta R(H, LJ)$ between Higgs boson candidate and leading b-tagged jet.
\end{enumerate}

\begin{figure}[]
  \centering
  \includegraphics[width=0.48\columnwidth]{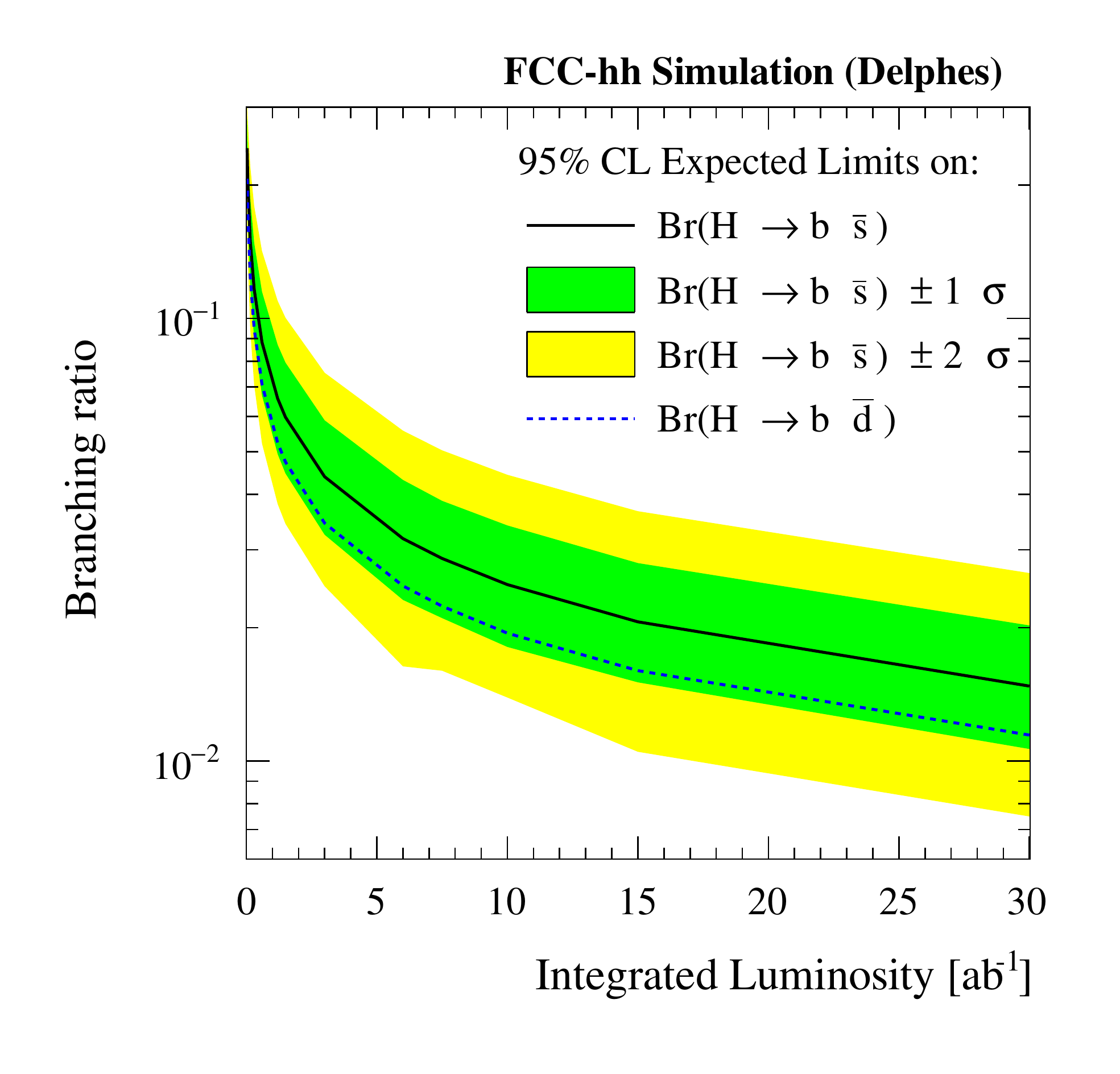}
  \includegraphics[width=0.48\columnwidth]{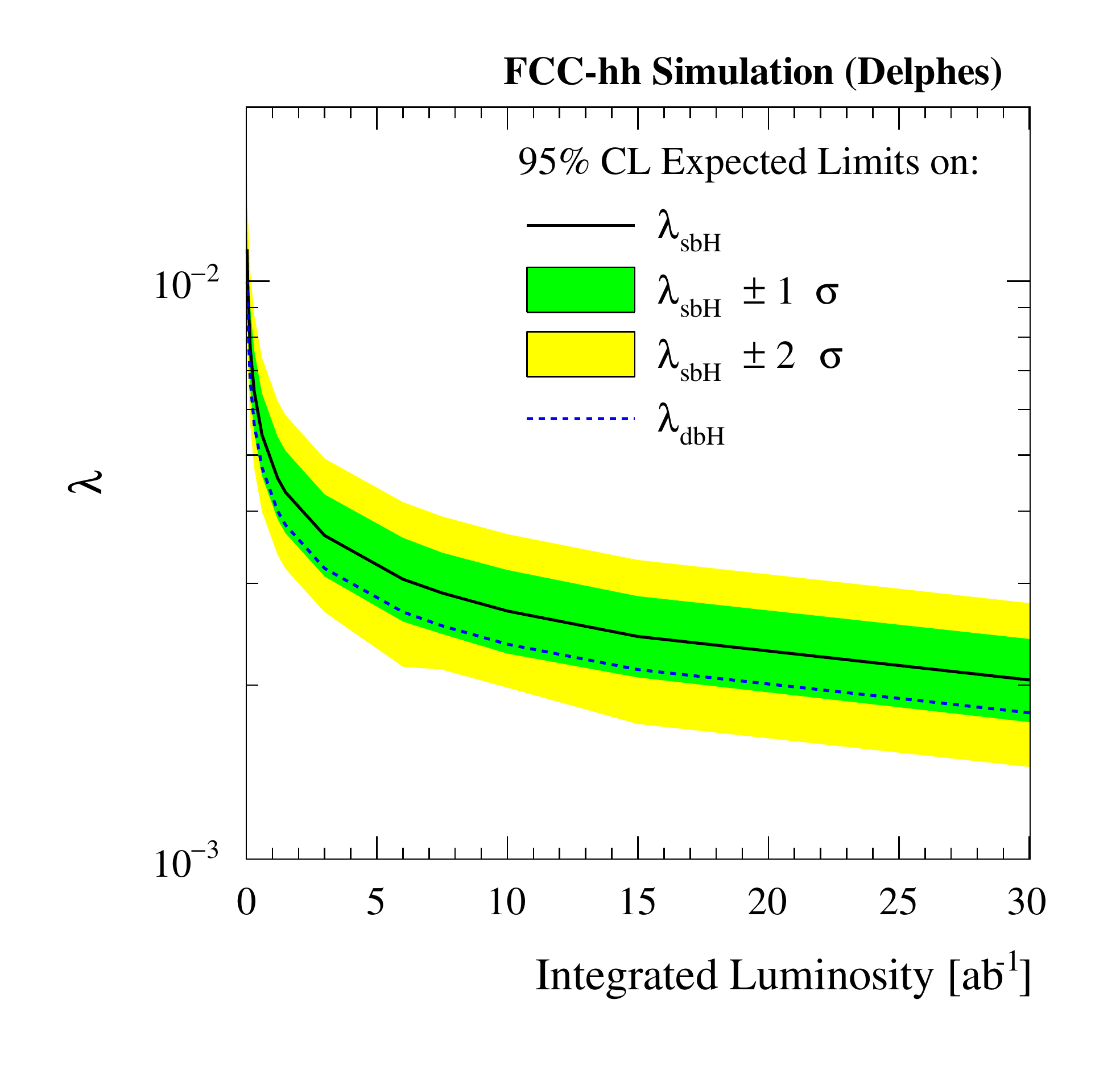}
  \caption{ 
    Expected exclusion limits at 95\% C.L. on the FCNC $H \rightarrow b\bar{s}$ and $H \rightarrow b\bar{d}$ branching fractions (left) and FCNC couplings (right) as a function of integrated luminosity. }
  \label{fig:fcc_plots_lumi}
\end{figure}

For each background a 20\% normalisation uncertainty is assumed and
incorporated in statistical model as nuisance parameter.  The asymptotic frequentist formulae
\cite{Cowan:2010js} is used to obtain an expected upper limit on signal cross section based on an Asimov data
set of background-only model. The 95\% C.L. expected exclusion limits on the branching fractions are given in Table \ref{FCC_limits}.
Figure \ref{fig:fcc_plots_lumi} shows the expected exclusion limits at 95\% C.L. on the FCNC $H \rightarrow b\bar{s}$ and $H \rightarrow b\bar{d}$ branching fractions and FCNC couplings as a function of integrated luminosity.

\begin{table}[]
      \centering
      \caption{ The 95\% C.L. expected exclusion limits at FCC-hh on the branching fractions of Higgs FCNC decays and flavor-violating couplings in comparison with present experimental limits. }
      \label{FCC_limits}
      \vspace{5pt}
      \renewcommand{\arraystretch}{1.5}
      \begin{tabular}{c|c c}
      \hline
      \hline
      Experiment                                                           & $\mathcal{B}(H \rightarrow b\bar{s})$ & $\mathcal{B}(H \rightarrow b\bar{d})$ \\ \hline
      Meson oscillations \cite{Harnik:2012pb}                              & $7 \times 10^{-3}$  & $8 \times 10^{-5}$     \\
      CMS LHC $H \rightarrow ZZ \rightarrow 4\ell$ (35.9 fb$^{-1}$, 13 TeV)& $27 \times 10^{-2}$ & $32 \times 10^{-2}$ \\
      HL-LHC $H \rightarrow ZZ \rightarrow 4\ell$  (3 ab$^{-1}$, 14 TeV)   & $5.8 \times 10^{-2}$  & $6.0 \times 10^{-2}$  \\
      FCC-hh $H\rightarrow \gamma\gamma$  (30 ab$^{-1}$, 100 TeV)          & $1.5 \times 10^{-2}$ & $1.1 \times 10^{-2}$ \\
      \hline
      \hline
      Experiment                                                           & $\lambda_{sbH}$      & $\lambda_{dbH}$      \\ \hline
      Meson oscillations \cite{Harnik:2012pb}                              & $1.9 \times 10^{-3}$ & $2.1 \times 10^{-4}$ \\
      CMS LHC $H \rightarrow ZZ \rightarrow 4\ell$ (35.9 fb$^{-1}$, 13 TeV)& $13 \times 10^{-3}$  & $14 \times 10^{-3}$  \\
      HL-LHC $H \rightarrow ZZ \rightarrow 4\ell$  (3 ab$^{-1}$, 14 TeV)   & $4.0 \times 10^{-3}$  & $4.1 \times 10^{-3}$  \\
      FCC-hh $H\rightarrow \gamma\gamma$ (30 ab$^{-1}$, 100 TeV)           & $2.0 \times 10^{-3}$ & $1.8 \times 10^{-3}$ \\
      \hline
      \end{tabular}
    \end{table}
    
\section{Conclusions}
In this work, we demonstrate that the contribution of flavour violation interaction to the production of the Higgs boson in high energy proton-proton collisions can be used 
for the direct search.
The realistic detector simulation and accurately reproducing analysis selections of the CMS Higgs boson measurement in the four-lepton final state at $\sqrt{s} = 13$ TeV
allow to set upper limits on the branching fractions of $H \to b\bar{s}$ and $H \to b\bar{d}$ and project the searches into HL-LHC conditions.
We also examine the sensitivity at FCC-hh based on Higgs boson production with $H \rightarrow \gamma \gamma$ decay channel.
Expected upper limits of the order of $10^{-2}$ at $95\%$ CL for $\mathcal{B}(H \rightarrow b\bar{s})$ and $\mathcal{B}(H \rightarrow b\bar{d})$
are competitive with the indirect limits from meson oscillations experiments.
The outcome of our study is summarised in Table \ref{FCC_limits}. Further improvements are possible through the combination of results of different Higgs boson decay and interaction searhes such as pair Higgs boson production. 

\section*{Acknowledgments}
We are grateful to H.~Gray‎, F.~Moortgat and M.~Selvaggi for permission to use the MC samples with background processes used in FCC-hh sensitivity study. 
We also would like to thank V.F.Kachanov, A.M.Zaitesv for useful discussions.



\bibliographystyle{elsarticle-num} 
\biboptions{numbers,sort&compress}
\bibliography{HiggsCouplings.bib}





\end{document}